\documentclass[12pt]{amsart}
\usepackage{geometry}                
\usepackage{graphicx}
\usepackage{amssymb}
\usepackage{amsmath}
\usepackage{amsthm}

\usepackage{mathtools}

\theoremstyle{definition}

\newtheorem*{thm*}{Theorem}

\newtheorem*{lem*}{Lemma}

\newcommand{\beq}{
\begin{eqnarray}
}
\newcommand{\eeq}{
\end{eqnarray}
}

\usepackage{epstopdf}
\title{Confinement for all couplings in a ${\mathbb Z}_{2}$ 
lattice gauge theory}

\author[Orland]{P. Orland}
\address{Department of Natural Sciences, Baruch College, CUNY, New York, NY, 10010-5585 and Physics Program, CUNY Graduate Center, New York, NY, 10016-4309}
\email{orland@nbi.dk}

\date{\today}                                           

\begin{document}

\begin{abstract}
For a particular lattice gauge theory with ${\mathbb Z}_2$ gauge invariance there is confinement for all 
couplings. The gauge fields, on lattice links, lie in the closed interval $[-1,1]$. It is proved that the expectation value of
a gauge-invariant loop operator decays as the exponential of minus the area. \end{abstract}

\maketitle

\section{Introduction}

A proof of quark confinement for the SU($3$) non-Abelian gauge theory in four spacetime dimensions has consistently eluded
all who have attempted it. The lattice approach is the clearest mathematically, but the depth of understanding of
lattice expectation values is too shallow to be the basis of a genuine proof. In this paper, we attempt to shine 
some light on this difficult subject by presenting a proof of confinement for a well-defined lattice gauge theory. The model
we investigate is peculiar in some respects:\begin{itemize}
\item The gauge group is ${\mathbb Z}_{2}$ (thus there are two colors), instead of a non-Abelian Lie group. 
\item The link variables are not representations of group elements.
\item Reflection positivity is not yet proved (and may actually be false).
\item Confinement holds for any dimension $d$, not just for $d\le 4$.
\end{itemize}
Nonetheless, the proof is so simple that we feel presenting it to be instructive.

The $d$-dimensional hypercubic lattice $\Lambda$ consists of sites $x=(x^{0},\dots,x^{d-1})$, which are $d$-tuples of integers. Our discussion works equally well for a lattice with periodic or open boundary conditions. We
denote spacetime directions by the indices $\mu$, $\nu$, etc. taking the values $0,\dots,d-1$. The domain of coordinates is
$x^{\mu}=1,2,\dots,L^{\mu}$, $\mu=0,\dots,d-1$.
We define unit
vectors ${\hat 0}=(1,0,0\dots,0)$, ${\hat 1}=(0,1,0\dots,0),\dots,{\widehat{d-1}}=(0,0,0\dots,0,1)$. We generally use
the letter $l$ to denote
such a link. A plaquette $P$ on this lattice is an elementary square of side unity. We write 
$x\in \Lambda$, $l\in\Lambda$ and $P\in\Lambda$ to indicate 
that the site $x$, the link $l$ and the plaquette $P$, respectively, belong to the lattice.

We assign the lattice ``spin" $\phi(l)\in [-1,1]$ to the link $l$. This lattice link field can be used to write a 
$\mathbb Z_{2}$-invariant gauge theory, with partition function
\begin{eqnarray}
Z=\left[\prod_{l\in \Lambda} \int_{-1}^{1} d\phi(l) \right] e^{-\beta S}, \;\;S=-\sum_{P\in \Lambda} \phi(P)+
\omega \sum_{l\in\Lambda}\phi(l)^{2}\,, \label{partfunct}
\end{eqnarray}
where $\beta$ and $\omega$ are nonnegative constants and 
\begin{eqnarray}
\phi(P)=\prod_{l\in P}\phi(l). \label{prod-plaquette}
\end{eqnarray}

The partition function (\ref{partfunct}) is invariant under the local ${\mathbb Z}_{2}$ gauge symmetry $\phi(l)\rightarrow \sigma(x) \phi(l) \sigma(x+{\hat \mu})$, where $l$ joins $x$ to $x+{\hat \mu}$ and where the gauge transformation is
$\sigma(x)=\pm 1$. What is also invariant is the Wilson loop, consisting of a closed contour $C$ of links $l\in C$, by
\begin{eqnarray}
A(C)=\prod_{l\in C}\phi(l)\,. \label{Wilson}
\end{eqnarray}
The expectation value of any function $F(\phi)$ of gauge fields $\phi(l)$ links $l$ is
\begin{eqnarray}
\langle F \rangle=Z^{-1}\left[\prod_{l\in \Lambda} \int_{-1}^{1} d\phi(l) e^{-\beta\omega \phi(l)^{2}}\right] F(\phi) 
\exp{\beta \sum_{P\in \Lambda} \phi(P)}.\label{expWilson}
\end{eqnarray}
The expectation value of 
$A(C)$ is the order parameter $\langle A(C)\rangle$. The lattice gauge theory is in a confined phase, if the order parameter falls off exponentially with the smallest possible area $\mathcal A$
of a surface whose boundary is $C$, that is $\langle A(C)\rangle \sim e^{-\sigma{\mathcal A}}$, where $\sigma$ is
the string tension. For $\omega=d-1$, the lower bound on the string tension vanishes in the limit
of zero gauge coupling. We do not know whether the true string tension also vanishes in that limit.

For the conventional ${\mathbb Z}_{2}$ lattice gauge theory, with gauge fields $\phi(l)=\pm 1$, the quantity $\phi(l)^{2}$
is trivially unity. In our model, however, this is a gauge-invariant quantity in the interval $[0,1]$.

In the light of the  remark of the previous paragraph, we point out that another loop order parameter of interest can be defined for this model. The sign of each link variable 
$\tau(l)={\rm sgn} \phi(l)=\pm1$ transforms the same way as $\phi(l)$ under a gauge transformation. In other words, $\tau(l)\rightarrow \sigma(x) \tau(l) \sigma(x+{\hat \mu})$, where again $l$ joins $x$ to $x+{\hat \mu}$. The variable $\tau(l)$ is similar to an
Ising gauge field. The quantity depending on the loop $C$, 
\begin{eqnarray}
A^{\prime}(C)=\prod_{l\in C}\tau(l)\in {\mathbb Z}_{2}\,, \label{IsingWilson}
\end{eqnarray}
is gauge invariant.


We next compare our lattice gauge model to the SU($2$) gauge theory with no matter. In the latter, 
we have link variables $U(l)$, lying in the fundamental unitary representation of SU($2$). If $l$ joins $x$ to $x+{\hat \mu}$, we may
write $U(l)=U_{\mu}(l)$. On the oriented plaquette with corner sites in the order (prescribed by the orientation) $x$, $x+{\hat \mu}$, $x+{\hat \mu}+{\hat \nu}$, $x+{\hat \nu}$, we assign the quantity 
\begin{eqnarray}
U(P)=U_{\mu}(x)U_{\mu}(x+{\hat \mu})U_{\nu} (x+{\hat \nu})^{-1}U_{\nu}(x)^{-1}. \nonumber
\end{eqnarray}
The SU($2$) gauge theory has partition function
\begin{eqnarray}
Z_{{\rm SU}(2)}=\left[\prod_{l\in \Lambda} \int_{{\rm SU}(2)} dU(l) \right]\;\exp \beta\sum_{P\in \Lambda} {\rm Tr}\, U(P), \nonumber
\end{eqnarray}
where the integration over each link is the Haar measure on SU($2$). Now the SU($2$) link variables may be written uniquely as product
$U(l)=\tau(l)V(l)$ of an element $\tau(l)$ of the center group ${\mathbb Z}_{2}$ and an element $V(l)$ of the coset group 
${\rm SU}(2)/{\mathbb Z}_{2}$. Then we may write the partition function \cite{MackPetKorth} as 
\begin{eqnarray}
Z_{{\rm SU}(2)}=\left[\prod_{l\in \Lambda}\int_{{\rm SU}(2)/{\mathbb Z}_{2}} dV(l)  \sum_{\tau\in {\mathbb Z}_{2}}\,\right]\;\exp \beta\sum_{P\in \Lambda} {\rm Tr}\, V(P) \tau(P). \nonumber
\end{eqnarray} 
The physical interpretation of writing the partition function this way, is that the variables in ${\rm SU}(2)/{\mathbb Z}_{2}$ {\em disorder}
those in ${\mathbb Z}_{2}$, making correlation functions, {\em e.g.}, loop expectation values, fall off faster with separation. One can
think of the SU($2$) gauge theory as a ${\mathbb Z}_{2}$ gauge theory with fluctuating coupling
$\beta \,{\rm Tr}\, V(P)$. Such an interpretation holds for our model
as well.

We may decompose $\phi(l)=\tau(l)f(l)$, where $\tau(l)\in {\mathbb Z}_{2}$ and $f(l)\in [0,1]$. The partition function (\ref{partfunct}) may
be written
\begin{eqnarray}
Z=\left[\prod_{l\in \Lambda} \int_{0}^{1} df(l) \sum_{\tau\in {\mathbb Z}_{2}}\, \right] e^{-\beta S}, \;\;S=-\sum_{P\in \Lambda} f(P)\tau(P)+
\omega \sum_{l\in\Lambda}f(l)^{2}\,. \nonumber
\end{eqnarray}
Here we see that the quantities on plaquettes $f(P)$ produce a fluctuating coupling $\beta \,f(P)$, disordering the
${\mathbb Z}_{2}$ degrees of freedom, similar to the interpretation of the SU$2$) theory above.

In the light of the above remarks, we note that large $\omega$ forces $f(l)$ close to zero, thus strongly-coupling the  ${\mathbb Z}_{2}$
degrees of freedom $\tau(l)$. In the theorem we present shortly, the reader should be able to see explicitly that the lower bound
on the string tension is not valid for arbitrarily small $\omega$.

The partition function (\ref{partfunct}) and the order parameter resemble expressions discussed by Weingarten
\cite{weingarten}. Weingarten described a lattice path-integral formulation of Nambu-Goto strings (he primarily discussed the case of
a complex lattice field, but also mentioned real fields. The complex and real cases correspond 
to oriented and unoriented strings, respectively). Replacing the integration in our model 
$\int_{-1}^{1}d\phi(l)$ by 
$\int_{-\infty}^{\infty}d\phi(l)$ yields the real version of Weingarten's partition function and order parameter. The model with unbounded measure does not actually exist; the Hamiltonian/action is quartic, producing an instability 
rendering the partition function and expectation values 
nonexistent (for either the real or complex cases) \cite{weingarten}. In our case, 
the finite domain of integration makes the model (\ref{partfunct}) well-defined.

The result of this paper is the
\begin{thm*}The order parameter for a closed rectangular contour $C$ is bounded above by 
\begin{eqnarray}
\langle A(C)\rangle \le Be^{-{\widetilde\sigma} {\mathcal A}}\,, \label{confinement!}
\end{eqnarray}
where $B$ is a constant, ${\mathcal A}$ is the area of the rectangle enclosed by the contour C and
\beq{\widetilde \sigma}=2\sinh^{-1}\frac{1}{\sqrt{8}}\left[\sqrt{\frac{4\omega}{\pi\beta}}e^{-\omega\beta-1}+2(\omega-d+1)\right], 
\label{str-tens}
\eeq
provided the right-hand side is positive.\end{thm*}

This theorem indicates confinement (of quarks of two colors) with string tension $\sigma\ge {\widetilde \sigma}$ for any choice of 
inverse coupling $\beta$. Notice that the requirement of positivity in (\ref{str-tens}) holds if the damping coefficient $\omega\ge d-1$. Confinement holds
independently of the dimension $d$; this is not what we expect of SU($N$) lattice gauge theories, which are believed to confine only for $d\le 4$. Nonetheless, we believe the result is interesting, because the dependence of the lower bound $\sigma$ on $\beta$ resembles asymptotically-free
behavior. We prove the theorem in Section 3.

We are convinced, but have not rigorously proved, that the expectation value of the loop expression in (\ref{IsingWilson}) also decays
exponentially with its area:
\begin{eqnarray}
\langle A^{\prime}(C)\rangle \le B^{\prime}e^{-{\widetilde\sigma} {\mathcal A}}\,. \label{IsingConfinement!}
\end{eqnarray}
We explain why we believe this to be true, after proving the theorem in Section 3.

As of this writing, we do not know whether the continuum limit exists as $\beta\rightarrow \infty$ or, in particular, whether $\sigma$ or the mass gap $M$ vanishes or how the ratio $M^{2}/\sigma$ behaves in this limit.

Reflection positivity, through hypersurfaces bisecting links, may not hold for this model. Unlike conventional lattice gauge theories for which the link variables are group 
elements, {\em e.g.}, in ${\mathbb Z}_{N}$, U($1$) or SU($N$), the variable $\phi(l)$ is not in ${\mathbb Z}_{2}$. Thus the action of the gauge group is not free (in the sense
of fiber bundles), meaning that the variables $\phi(l)$ cannot be fixed, except possibly up to a sign. We say more about this in the last section
of this paper.

In the next section, we prove a lemma concerning the presence of a mass gap in a spin model. This property is used
in Section 3 to prove the theorem, where we give our arguments for (\ref{IsingConfinement!}). In the last section, we discuss how our results may be extended and, in particular, the issue of reflection positivity. 


\section{Surfaces between two walls}\label{sec:properties}

In this section, we establish a preliminary lemma, which we will use to prove the theorem. 

The proofs of the lemma and the theorem rely on standard Griffiths-Kelley-Sherman (GKS) inequalities \cite{ginibre}. Consider an
arbitrary ferromagnetic Hamiltonian on a lattice $\Lambda$, with variables $\phi(q)$ on elementary locations $q$ (which can be sites, links, plaquettes, etc.)
\beq
H(\phi)=-\sum_{A\subset \Lambda} J_{A}\phi^{A},\;J_{A}\ge 0,\;\; \phi^{A}=\prod_{q\in A} \phi(q) \nonumber
\eeq
with expectation value of a quantity $F(\phi)$ and partition function, respectively,
\beq
\langle F\rangle=Z^{-1}\int \left[\prod_{q\in \Lambda} dm(\phi)\right] F(\phi)e^{-H(\phi)},\;\;Z=\int\left[\prod_{q\in \Lambda} dm(\phi)\right]
e^{-H(\phi)}, \nonumber
\eeq
with positive
measure $dm(\phi)$ over $\mathbb R$, satidfying $dm(-\phi)=dm(\phi)$. Then
$\langle \phi^{A}\rangle\ge 0,$ for $A\subset \Lambda$ (GKS I) and 
$\langle \phi^{A}\phi^{B}\rangle \ge \langle \phi^{A}\rangle\langle\phi^{B}\rangle,$ for $A,B\subset \Lambda$ (GKS II).

It is useful to 
generalize the measure over $[-1,1]$, to a measure over the entire real line, which depends on a positive integer $p$ \cite{McB+S}:
\begin{eqnarray}
dm_{p}(\phi)= d\phi \,e^{-\beta\omega \phi(l)^{2}-\phi^{2p}}. \label{smoothmeasure}
\end{eqnarray}
We recover a Gaussian measure on the interval $[-1,1]$ in the limit of large $p$:
\begin{eqnarray}
\int_{-1}^{1}d\phi(l)\, e^{-\beta\omega \phi(l)^{2}}
=\lim_{p\rightarrow \infty} \int_{\mathbb R} dm_{p}[\phi(l)].    \nonumber 
\end{eqnarray}

It has long been known that there is a mass gap in a spin model describing the statistical mechanics of a (d-1)-dimensional
membrane between two (d-1) dimensional parallel hard walls. McBryan and Spencer proved that a version of such a model has exponentially-decaying correlation functions for any coupling \cite{McB+S}. Their results were improved upon by Bricmont, El Mellouki and Fr\"{o}hlich \cite{BMF}.

The two-wall model on  a square $k$-dimensional hypercubic lattice $\Lambda$, has the partition function and correlation function, respectively,
\beq
Z=\left[\prod_{y} \int_{-1}^{1}d\phi(y) e^{-\beta\omega \phi(y)^{2}}\right]\;e^{\beta\sum_{y,\mu}  \phi(y)\phi(y+{\hat \mu}  )}, \label{MBSZ}
\eeq
and
\beq
\langle \phi(0)\phi(x)\rangle=Z^{-1}\left[\prod_{y} \int_{-1}^{1}d\phi(y) e^{-\beta\omega \phi(y)^{2}}\right] \phi(0)\phi(x)
e^{\beta\sum_{y,\mu}  \phi(y)\phi(y+{\hat \mu}  )}, \label{MBSCorr}
\eeq
where $x$ refers to the sites of the lattice and where $\mu=1,\dots,k$. 

\begin{lem*} If
\beq
{\widetilde s}=\sqrt{\frac{4\omega}{\pi\beta}}e^{-\omega\beta-1}+2(\omega-k)>0, \label{pos-s}
\eeq
then the correlation function (\ref{MBSCorr}) decays exponentially,
\beq
\langle \phi(0)\phi(x)\rangle\le b e^{-{\widetilde m}\vert x\vert}, \nonumber
\eeq
where $\widetilde m$ (not to be confused with the measure (\ref{smoothmeasure})) is
\beq
{\widetilde m}=2\sinh^{-1}\sqrt{ \frac{{\widetilde s}^{\,2}}{8}},\nonumber
\eeq
and $b$ is a constant.
\end{lem*}

We briefly explain how the expressions (\ref{MBSZ}) and (\ref{MBSCorr}) can be understood in terms of a membrane
between two hard walls separated by $2D$. The partition
function for the latter is
\beq
Z=\left[\prod_{y} \int_{-D}^{D}d\phi(y) \right]\;e^{-\frac{1}{2} \sum_{y,\mu} [\phi(y+{\hat \mu})-\phi(y)]^{2}-\frac{r}{2}\sum_{y}\phi(y)^{2}}. \nonumber
\eeq
Rescaling $\phi(x)$ by a factor of $1/D$ yields (\ref{MBSZ}), with $\beta=D^{2}$, $\omega=k+r/2$. In the earlier studies \cite{McB+S}, \cite{BMF}, $r=0$.

Later we will set $k=d-1$, where $d$ is the dimension of the gauge theory 
lattice. We present the proof, which closely resembles McBryan and Spencer's, only because 
we are imposing (\ref{pos-s}), which is a weaker condition 
than
$\omega=k$ (assumed in References \cite{McB+S} and \cite{BMF}). 

\begin{proof}
We first replace the measure by $dm_{p}(\phi)$ in (\ref{MBSCorr}) and (\ref{MBSZ}), and extend the range of integration over the entire real line. The Dyson-Schwinger equation
\beq
&{\bigg[}{\prod}_{y} \int_{-\infty}^{\infty} d\phi(y){\bigg ]}& \!\!\!\frac{\partial}{\partial \phi(x)} {\bigg(} \phi(0)  \nonumber \\
&\times& \!\!\!\!\!\!\!\!\!\!\!\!\!\! \!\!\!\!\!\!\!\!\!  \left. \exp\left\{-\sum_{y} [ \beta\omega \phi(y)^{2}+\phi(y)^{2p} ]+\beta\sum_{y,\mu}  \phi(y)\phi(y+{\hat \mu}) 
 \right\} \right)=0,
\nonumber
\eeq
follows from elementary calculus. We write this as
\beq
0\!\!&\!\!<\!\!&\!\!\beta (-\Delta+s)   \langle \phi(0) \phi(x)\rangle \nonumber \\
&=&\delta_{0,x}-  2p\langle \phi(0) \phi(x)^{2p-1} \rangle
+\left[(s-2\omega+2k)\beta \right]  \langle \phi(0) \phi(x)\rangle],  \label{SDeq}
\eeq
for arbitrary positive $s$, where $\Delta$ is the lattice Laplacian, whose action on a function $f(x)$ is
\beq
\Delta f(x)=\sum_{\pm\mu}[ f(x\pm \mu)-f(x)], \label{Lapl}
\eeq
and the inequality on the left-hand side of (\ref{SDeq}) follows from GKS I and positive 
semi-definiteness of the operator (\ref{Lapl}), for standard (for example periodic or open) boundary conditions. By GKS II, 
\beq
2p\langle \phi(0) \phi(x)^{2p-1} \rangle\ge \langle \phi(0) \phi(x) \rangle\;2p \langle \phi(x)^{2p-2} \rangle \label{an-ineq},
\eeq
and GKS II again, 
\beq
2p\langle \phi(x)^{2p-2}\rangle &\ge&  2p \left\{ \int_{-\infty}^{\infty} d\phi(x) \exp[ -\omega\beta \phi(x)^{2}-\phi(x)^{2p}] \right\}^{-1}
\nonumber\\
&\times&\int_{-\infty}^{\infty} d\phi(x) \exp[ -\omega\beta \phi(x)^{2}-\phi(x)^{2p}] \;\phi(x)^{2p-2} \nonumber\\
&\ge& 2p\frac{\int_{-1}^{1} d\phi\, e^{-\omega\beta\phi^{2}}e^{-\phi^{2p}}   \phi^{2p-2} }{ \int_{-\infty}^{\infty} d\phi\, e^{-\omega\beta \phi^{2}}}
\ge 2p\frac{\int_{-1}^{1} d\phi\, e^{-\omega\beta-1}  \phi^{2p-2} }{ \sqrt{\frac{\pi}{\omega\beta} } }\nonumber \\
&=&\frac{2p}{2p-1}\,2\sqrt{\frac{\omega\beta}{\pi}}e^{-\omega\beta-1} \;\stackrel{-\!-\!-\!-\!-\!\longrightarrow}{p\rightarrow \infty}\;
\sqrt{\frac{4\omega\beta}{\pi}}e^{-\omega\beta-1} .\label{another-ineq}
\eeq
Substituting (\ref{an-ineq}) and (\ref{another-ineq}) into (\ref{SDeq}),
\beq
0&<&\beta (-\Delta+s)   \langle \phi(0) \phi(x)\rangle \nonumber \\
&\le& \delta_{0,x} +\beta \left[s-2(\omega-k)-\sqrt{\frac{4\omega}{\pi\beta}}e^{-\omega\beta-1}\right]  \langle \phi(0) \phi(x)\rangle\,.
\nonumber
\eeq
Thus
\beq
0<\beta (-\Delta+s)   \langle \phi(0) \phi(x)\rangle \le \delta_{0,x}\,,\nonumber
\eeq
for any choice of $s\le {\widetilde s}$. Since $\langle \phi(0) \phi(x)\rangle\ge 0$, by GKS I and since $-\Delta+s$ is positive and nonsingular,  we have 
\beq
\langle \phi(0) \phi(x)\rangle\le \frac{1}{\beta(-\Delta+{\widetilde s})} \,\,\delta_{0,x}\,,\nonumber
\eeq
which completes the proof.
\end{proof}

If we set $\omega=k$ \cite{McB+S}, \cite{BMF} and define the coupling $g=\beta^{-1}$, we 
have ${\widetilde s}=\sqrt{\frac{4kg}{\pi}}e^{-k/g-1}$, which is similar to the scaling of dimensionful quantities in asymptotically-free field theories. It was proved in Reference \cite {BMF} that the dependence of the gap 
on $\beta$, for $\omega=k$ is indeed of this form, for $k\ge 3$. For
$k=1,2$, the behavior is different. We will not comment on this further, because $k=3$ 
corresponds to the four-dimensional gauge theory, which we feel is the most interesting case.

\section{Proof of Confinement}\label{sec:Proof}

The essential idea used in the proof below is to view the gauge theory in $d$ dimensions as a set of coupled $(d-1)$-dimensional spin models \cite{DurFr}. The same observation was independently made in the Hamiltonian 
framework \cite{OCC}, where it was used to compute string tensions and mass 
gaps for some weakly coupled SU($N$) gauge theories in 2+1 dimensions.

\begin{proof} [Proof of the theorem] We modify (\ref{partfunct}) by changing the coefficient of $\sum_{P} \phi(P)$ to $\tilde \beta$ at plaquettes in planes parallel to
$x^{j}$ and $x^{k}$, $j,k=1,\dots,d-1$. The coefficients of the other plaquettes, in the planes parallel to $x^{0}$ and $x^{j}$, $j=1,\dots,d-1$ are not
modified, nor is the quadratic term on links. The exponent in the partition function is changed by
\beq
\beta S\rightarrow -{\widetilde \beta} \sum_{P\,\parallel\, (x^{j},x^{k})}\left[ \phi(P)-1\right] -\beta \sum_{P\,\parallel \,(x^{0},x^{j})} \phi(P)+
\beta \omega \sum_{l\in\Lambda}\phi(l)^{2}, \nonumber
\eeq
where we have subtracted an irrelevant constant. If we take ${\widetilde \beta}\rightarrow \infty$, then $\phi(P)$ on every plaquette $P$ parallel to $(j,k)$ planes is forced to be unity, and every space-like link variable is forced to be $\phi(l)=s(x)s(x+{\hat j})$, $s(x)=\pm1$, where $l$ joins
$x$ to $x+{\hat j}$, $j\neq 0$. Note that the value of $\phi(l)$ on a link $l$ joining $x$ to $x+{\hat 0}$ is not similarly constrained.

We write $\phi(l)=\phi_{0}(x^{0}, x_{s})$, where $x_{s}=(x^{1},\dots,x^{d-1})$, and $l$ joins $x$ to $(x^{0}+1, x_{s})$. We redefine
$\phi_{0}(x^{0}, x_{s})\rightarrow s(x^{0},x_{s}) \phi_{0}(x^{0}, x_{s})s(x^{0}+1,x_{s})$. Then
the partition function factorizes into 
\beq
Z=\prod_{x^{0}} Z_{x^{0}}\,, \;
Z_{x^{0}} \!\!&\!\!=\!\!&\!\! \left[\prod_{x_{s}}\int_{-1}^{1} d\phi_{0}(x^{0},x_{s})e^{-\beta\omega \phi_{0}(x^{0},x_{s})^{2}}\right] \;  e^{-\beta S_{x^{0}}},\nonumber\\
S_{x^{0}}\!\!&\!\!=\!\!&\!\!-\sum_{x_{s}}\sum_{j=1}^{d-1}\beta \phi_{0}(x^{0},x_{s})\phi_{0}(x^{0},x_{s}+{\hat j})\,. \nonumber
\eeq	
Each factor $Z_{x^{0}}$ is a partition function of the model in the lemma. 

The expectation value of a square Wilson loop, whose corners are at \begin{itemize}
\item $x^{0}=0, x_{s}=0$, 
\item $x^{0}=0, x_{s}=N_{1}{\hat 1}$,
\item $x^{0}=N_{0}, x_{s}=0$, 
\item $x^{0}=N_{0}, x_{s}=N_{1}{\hat 1}$,
\end{itemize}
is 
\beq
\left\langle 
\prod_{l\in N_{0}\times N_{1}} \phi(l)\right\rangle
=\langle \phi(0)\phi(N_{1}{\hat 1})\rangle^{N_{0}},   \nonumber
\eeq
where each of the expectation values on the right-hand side is with respect to a hard-wall model, with dimension
$k=d-1$. Thus, by the lemma,
\beq
\left\langle 
\prod_{l\in N_{0}\times N_{1}} \phi(l)\right\rangle \le
b^{N_{0}}e^{-{\widetilde m}N_{0}N_{1}}, \label{area-law-decay}
\eeq
which is an area decay law.

By GKS II, the loop on the left-hand side of (\ref{area-law-decay}) falls off at least as fast as the right-hand side, for finite 
$\tilde \beta$. In particular, this is true if
${\tilde \beta}=\beta$, establishing (\ref{confinement!}).
\end{proof}

We conclude this section with our reasoning behind a claim we made in Section 1, that a loop order parameter with purely Ising type link
variables decays exponentially with its area, {\em i.e.},
(\ref{IsingConfinement!}). Consider, for the model of two hard walls (\ref{MBSZ}), the correlation function of sign variables $\tau(x)={\rm sgn}\phi(x)$,  that is
$\langle \tau(0) \tau(x)\rangle$. We expect that this correlation function decays exponentially with $\vert x\vert$, in the same way
as $\langle \phi(0) \phi(x)\rangle$. Were this not the case, massless spin waves would exist in the 
parallel-hard-wall model of Section 2. Then the justification of  the area law of the Ising order 
parameter works exactly the same way as the proof of the theorem. A weaker result, which is 
easier to prove, is that the order parameter of the standard Ising gauge theory is an upper 
bound on both $\langle A(C)\rangle$ and on $\langle A^{\prime}(C)\rangle$
in our model, by applying GKS II, as done for SU($N$) gauge theories in Reference \cite{MackPetKorth}.

\section{Discussion}\label{sec:Disc}

To summarize, we have introduced a lattice gauge model with ${\mathbb Z}_{2}$ gauge invariance, and shown that it confines for
all choices of inverse coupling $\beta$. The lattice gauge field $\phi(l)$ on a link $l$ does not lie within the gauge group, but the 
interval $[-1,1]$. The Wilson loop was defined as the product of lattice gauge fields around a contour. We argued, however, that
an alternative loop with a product of the sign of $\phi$ (that is, an Ising spin assigned to the link) also decays 
as the exponential of its area.

Our proof of both the lemma and main theorem depends heavily on the use of GKS inequalities. Such inequalities are not proved
for general ferromagnetic systems with non-Abelian symmetry. Some models for which they can be proved \cite{myGKS} are far too specific
to have broader application. Indeed, even for ${\mathbb Z}_{N}$, $N\ge 3$ systems, these inequalities are less applicable than for
the ${\mathbb Z}_{2}$ case. For spin systems with variables lying on a unit circle, such as classical XY models or U($1$) gauge theories,
GKS II is well-established \cite{ginibre}, but this is not so for variables of arbitrary norm. We hope to study some 
${\mathbb Z}_{N}$ or even SU($N$) gauge theories for which the action of the gauge group is not free,
for arbitrary $N$, but suspect this can only be done by a comparison with the ${\mathbb Z}_{2}$ model described above
(by, for example, establishing inequalities between the loop order parameters of the different systems). For all of these reasons, another proof of our theorem, which does not use GKS inequalities, would be handy. 

Our model is not obviously reflection positive (and indeed may not be) in the standard sense. It does satisfy a more restricted type of reflection positivity, where the reflections are through hypersurfaces containing spins $\phi(l)$ \cite{FILS}. The latter may be useful in investigating the model, but does not guarantee that the transfer matrix generates a contraction semigroup, hence a particle spectrum  \cite{OsterSeil}. 

We cannot investigate the question of reflection positivity by gauge fixing links (for example, in the temporal direction) used in the standard
proof of reflection positivity for lattice gauge theories \cite{OsterSeil}. In principle, however, no gauge fixing is needed. Consider the hyperplane on the lattice
$\Lambda$ halfway between between two parallel spacelike hypersurfaces of links. In other words, consider the locations with coordinates:
$(x^{0}+1/2, x^{1},x^{2},\dots,x^{d-1})$ as the points of the reflection plane (if periodic boundary conditions are taken, with even period $L^{0}$ in the temporal direction, we must
also include coordinates $(x^{0}+L^{0}/2+1/2, x^{1},x^{2},\dots,x^{d-1})$). Consider the link $l$ joining $x$ to $x+{\hat 0}$, and
split it into two, one half link $l^{\prime}$ joining $x$ to $x+{\hat 0}/2$ and a second half link $l^{\prime\prime}$ joining $x+{\hat 0}/2$ to
$x+{\hat 0}$. We place variables $\psi(l^{\prime})$ and $\psi(l^{\prime\prime})$ on these half links. We would have reflection positivity if there exists a measure $dM(\psi)$
on half-link variables such that
\beq
\int dM(\psi(l^{\prime}))^{*}dM(\psi(l^{\prime\prime}))\; \delta[\psi(l^{\prime}) \psi(l^{\prime\prime}) -\phi(l) ]
=\int dm(\phi(l))  \;.\label{ref-pos}
\eeq
where $dm(\phi(l))$ is the lattice-gauge-theory measure on links. It may be that $dM(\psi)$ does not exist for our model; but there
may a similar model where confinement can be proved, for which there does exist such a measure satisfying (\ref{ref-pos}). If so, it 
would be interesting
to study the Hilbert space of such a model further.


\bibliographystyle{plain}
\bibliography{ref}

\end{document}